\documentclass[conference]{IEEEtran}
\IEEEoverridecommandlockouts
\usepackage{bm}
\usepackage{cite}
\usepackage{amsmath,amssymb,amsfonts,steinmetz}
\usepackage{algorithmic}
\usepackage{graphicx}
\usepackage{textcomp}
\usepackage{xcolor}
\def\BibTeX{{\rm B\kern-.05em{\sc i\kern-.025em b}\kern-.08em
		T\kern-.1667em\lower.7ex\hbox{E}\kern-.125emX}}

\usepackage{pgfplots}
\usepackage[font=footnotesize]{caption}
\usepackage{subcaption}
\usepackage[ruled,vlined]{algorithm2e}
\usepackage{algorithmic}
\usepackage{bbm}
\usepackage{tikz}
\usetikzlibrary{calc}
\makeatletter
\newcommand{\gettikzxy}[3]{%
	\tikz@scan@one@point\pgfutil@firstofone#1\relax
	\edef#2{\the\pgf@x}%
	\edef#3{\the\pgf@y}%
}
\makeatother
\usepackage[draft]{todonotes}   % notes showed
\usepackage{upgreek} %use for mathbf symbols like \bm{theta} = \mathbf{\uptheta}
\usepackage{seqsplit}
\usepackage{hyperref}
\usepackage{balance}
\newcommand{\vind}[1]{\boldsymbol{#1}}
\newcommand{\matInd}[1]{\boldsymbol{#1}}
\newcommand{\coordinateMark}{v}
\newcommand{\risCoordinateVector}[1]{\vind{\coordinateMark}_{#1}}
\newcommand{\risPosition}{\vind{p}_{\risMark}}
\newcommand{\bsPosition}{\vind{p}_{\bsMark}}
\newcommand{\uePosition}{\vind{p}}
\newcommand{\risRotationMatrix}{\matInd{R}}
\newcommand{\rbuFont}[1]{\mathsf{#1}}
\newcommand{\risMark}{\rbuFont{r}}
\newcommand{\bsMark}{\rbuFont{b}}

\newcommand{\mr}[1]{\mathrm{#1}}
\newcommand{\defEq}{\triangleq}

\newcommand{\GainBs}{\gainMark_{\bsMark}}
\newcommand{\GainRis}{\gainMark_{\risMark}}
\newcommand{\Es}{E_{\mr{s}}}
\newcommand{\timeBias}{\Delta_\mr{t}}
\newcommand{\delayMark}{\tau}
\newcommand{\delayBS}{\delayMark_{\bsMark}}
\newcommand{\delayRis}{\delayMark_{\risMark}}
\newcommand{\aodRisToUeEl}{\aodMark_{\elevationMark}}
\newcommand{\aodRisToUeAz}{\aodMark_{\azimuthMark}}
\newcommand{\aodRisToUe}{\vind{\aodMark}}
\newcommand{\aodMark}{\phi}
\newcommand{\gainMark}{g}
\newcommand{\numSubCarriers}{N}
\newcommand{\oneVector}{\mathbf{1}}
\newcommand{\elementMultiply}{\odot}
\newcommand{\complexSet}{\mathbb{C}}
\newcommand{\deltaF}{\Delta f}

\newcommand{\risResponseVector}[1]{\vind{a}(#1)}
\newcommand{\aoaBsToRisEl}{\theta_{\mr{el}}}
\newcommand{\aoaBsToRisAz}{\theta_{\mr{az}}}
\newcommand{\aoaBsToRis}{\vind{\theta}}
\newcommand{\generalAngle}{\vind{\psi}}
\newcommand{\waveNumberVector}[1]{\vind{k}(#1)}

\newcommand{\risElementPosition}[1]{\vind{q}_{#1}}
\newcommand{\generalAngleAz}{\psi_{\mr{az}}}
\newcommand{\generalAngleEl}{\psi_{\mr{el}}}

\newcommand{\chParMark}{\mr{ch}}
\newcommand{\fisherInfMark}{\matInd{F}}
\newcommand{\positionOrientationMark}{\mr{po}}
\newcommand{\fisherInfpo}{\fisherInfMark_{\positionOrientationMark}}

\newcommand{\parameterMark}{\vind{\zeta}}
\newcommand{\parameterCh}{\parameterMark_{\chParMark}}
\DeclareMathOperator{\tr}{\mr{tr}}
\newcommand{\parameterPo}{\parameterMark_{\positionOrientationMark}}
\newcommand{\GainBsR}{\gainMark_{\bsMark,\mr{r}}}
\newcommand{\GainRisR}{\gainMark_{\risMark,\mr{r}}}
\newcommand{\GainBsI}{\gainMark_{\bsMark,\mr{i}}}
\newcommand{\GainRisI}{\gainMark_{\risMark,\mr{i}}}
\newcommand{\jacob}{\matInd{J}}
\newcommand{\fisherInfoch}{\fisherInfMark_{\chParMark}}
\newcommand{\realSet}{\mathbb{R}}
\newcommand{\rond}{\partial}
\newcommand{\azimuthMark}{\mr{az}}
\newcommand{\elevationMark}{\mr{el}}
\newcommand{\realPart}{\operatorname{Re}}
\newcommand{\vecNorm}[1]{{\left\lVert#1\right\rVert}}
\newcommand{\kron}{\otimes}
\DeclareMathOperator{\vect}{\mr{vec}}
\newcommand{\risToUeInRisCoorIndx}[1]{[{\vind{s}}]_{#1}}
\newcommand{\risToUeInRisCoor}{{\vind{s}}}
\DeclareMathOperator{\atant}{\mr{atan2}}
\newcommand{\meanSig}{\matInd{E}}

\newcommand{\lefto}{\mathopen{}\left}
\newcommand{\righto}{\right}

\newcommand{\herm}{{\textrm{H}}}

\newcommand{\ySumCol}{\vind{y}_{\mr{c}}}
\newcommand{\ySumColBar}{\bar{\vind{y}}_{\mr{c}}}

\newcommand{\rispVecT}[1]{\vind{a}_{\mr{c}}(#1)}
\newcommand{\rispVecH}[1]{\vind{a}_{\mr{r}}(#1)}

\newcommand{\betar}{\beta_{\mr{r}}}
\newcommand{\betac}{\beta_{\mr{c}}}
\newcommand{\Mrr}{M_{\mr{r}}}
\newcommand{\Mcc}{M_{\mr{c}}}

\newcommand{\FT}{\matInd{F}_{\mr{r}}}
\newcommand{\FH}{\matInd{F}_{\mr{c}}}
\newcommand{\NF}{N_{\mr{F}}}
\newcommand{\NFr}{N_{\mr{F}_{\mr{r}}}}
\newcommand{\NFc}{N_{\mr{F}_{\mr{c}}}}

\begin{document}

%\title{From Nothing to 3D Localization and Synchronization via One RIS}

%\title{RIS-Enabled Joint 3D Downlink Localization and Synchronization with Single-Antenna Base-Station and User-Equipment}
\title{SISO RIS-Enabled Joint 3D Downlink Localization and Synchronization}

\author{\IEEEauthorblockN{Kamran Keykhosravi\IEEEauthorrefmark{1}, Musa Furkan Keskin\IEEEauthorrefmark{1}, Gonzalo Seco-Granados\IEEEauthorrefmark{2}, and Henk Wymeersch\IEEEauthorrefmark{1}}
\IEEEauthorblockA{\IEEEauthorrefmark{1} Department of Electrical Engineering, Chalmers University of Technology, Sweden \\
\IEEEauthorrefmark{2}  Department of Telecommunications and
Systems Engineering, Universitat Auton\`{o}ma de Barcelona, Spain \\
email: kamrank@chalmers.se}

}

\maketitle

\begin{abstract}
We consider the problem of joint three-dimensional localization and synchronization for a single-input single-output (SISO) multi-carrier system in the presence of a reconfigurable intelligent surface (RIS), equipped with a uniform planar array. First, we derive the Cram\'er-Rao bounds (CRBs) on the estimation error of the channel parameters, namely,  the angle-of-departure (AOD), composed of azimuth and elevation, from RIS to the user equipment (UE) and times-of-arrival (TOAs) for the  path from the base station (BS) to UE and BS-RIS-UE reflection. In order to avoid high-dimensional search over the parameter space, we devise a low-complexity estimation algorithm that performs two 1D searches over the TOAs and one 2D search over the AODs. Simulation results demonstrate that the considered RIS-aided wireless system can provide submeter-level positioning and synchronization accuracy, materializing the positioning capability of Beyond 5G networks even with single-antenna BS and UE. Furthermore, the proposed estimator is shown to attain the CRB at a wide interval of distances between UE and RIS. Finally, we also investigate the scaling of the position error bound with the number of RIS elements.%}
\end{abstract}

\begin{IEEEkeywords}
Reconfigurable intelligent surface, position error bound,  Cram\'er-Rao bound, localization.
\end{IEEEkeywords}

\section{Introduction}
Provisioning of radio localization in cellular wireless communication standards was stirred by governmental requirements (Federal Communications
Commission and European Commission) on locating the user equipment (UE) by the operators upon receiving an emergency call. Other applications include location-aware communication, geo-advertising, augmented reality, etc \cite{surveyRosado}.  
According to 3GPP standards, localization in 4G   \cite{3gpp.36.855} can be obtained using time-difference-of-arrival measurements between a UE and multiple synchronized base stations (BSs). Since  UE is not synchronized to the BSs, at least 4 BSs are needed to solve for the 3D UE position and the clock bias. Hence, cellular localization requires sufficient dense infrastructure and tight inter-BS synchronization.
In the context of 5G, it has been shown that mmWave localization  with a single BS is possible, provided that UE and BS are equipped with large arrays, and communicate under sufficiently rich multipath conditions \cite{Shahmansoori18TWC}. By estimating angles-of-arrival (AOAs)  and angles-of-departure (AODs)  in addition to delays, localization and synchronization is possible, even when the multipath sources are a priori unknown \cite{wymeersch20185g}.
 Towards 6G, reconfigurable intelligent surfaces (RISs) deserve special attention among the technological enablers,  since they can redirect strong signals from a BS towards a UE, thus boosting communication quality among other advantages \cite{2019Basar}. In recent years, significant effort has been devoted to study  manufacturing, channel models, control, performance gains, and topology design of RIS-aided communication system (see e.g., \cite{dardari2020communicating,ICCRIScapacity}). 

 RISs have been considered as one of the main enabling technologies of beyond-5G localization \cite{bourdoux20206g,wymeersch2019radio} and have been investigated in a number of studies in the localization literature, e.g., \cite{guidi2019radio,abu2020near,he2020large,wang2020joint,wymeersch2020beyond,elzanaty2020reconfigurable}. In \cite{guidi2019radio,abu2020near}  localization in the near-field range of a RIS, functioning as a lens, is studied.
 Cram\'er Rao bounds (CRBs) have been derived in \cite{he2020large} for a 2D localization in presence of  a large intelligent surface equipped with a uniform linear arry (ULA). In \cite{wang2020joint}, authors have proposed a joint positioning and beam training algorithm for a wireless system comprising multiple RISs with ULAs. 
 A single-input single-output 2D localization problem with synchronized signaling and multiple RISs (with ULA) has been investigated in \cite{wymeersch2020beyond} by deriving the CRB bounds.
In \cite{elzanaty2020reconfigurable}, position error bounds (PEBs) have been derived for 3D uplink localization considering both synchronous and asynchronous cases. The authors pointed to the fact that the bounds can be attained by applying a grid search to maximize the log-likelihood function, however, no low-complexity estimation algorithm was presented. 

In this paper, we study the 3D downlink localization and synchronization of a UE with a single antenna using the received line-of-sight (LOS) signal from a single-antenna BS and the reflected signal from a RIS with a uniform planar array (UPA). To the best of our knowledge, RIS-aided localization has never been studied in this elemental configuration. The contributions of this paper are:

\begin{itemize}
    \item We show that both the 3D position and clock bias of the user can be estimated when both the BS and the UE have a single antenna, by the aid of a single RIS, provided that multiple distinct RIS phase profiles are used. As it is known, localization and synchronization would not be possible in the considered scenario without the RIS.
    %In the considered scenario in the absence of the RIS  the user position cannot be estimated, even in a 2D surface and the synchronous case. With the addition of 1 RIS, we show that both the use 3D position and clock bias can be recovered, provided that multiple distinct RIS phase profiles are used. 
    \item We perform the Fisher information matrix (FIM)  analysis and calculate the CRB bounds for the user position and channel parameters, showing that the localization is enabled by AOD estimation from the RIS.  
    \item We provide a novel low-complexity localization and synchronization method, which transforms the underlying 4-dimensional problem into two 1-dimensional and one two 2-dimensional problems. We also show that the proposed method attains the CRB bound. 
\end{itemize}
The MATLAB implementation of the CRB calculations and the proposed estimation technique can be found in the repository \cite{GithubRepo2020}.

\subsection*{Notation}
Vectors and matrices are indicated by lowercase and uppercase bold letters resp. The element in the $i$th row and $j$th column of matrix $\matInd{A}$ is specified by $[\matInd{A}]_{i,j}$. Similarly, $[\vind{p}]_i$ indicates the $i$th element of vector $\vind{p}$. The subindex $i:j$ specifies all the elements between $i$ and $j$. All vectors are columns, unless stated otherwise.  By $\kron$, we indicate the Kronecker product, and by $\vect(\cdot)$, the matrix vectorization operator. The complex conjugate, Hermitian, and transpose operators are represented by $(\cdot)^*$, $(\cdot)^\herm$, and $(\cdot)^\top$, respectively.

\section{System and Channel Model}

\subsection{System Setup}
\begin{figure}
    \centering
    \begin{tikzpicture}
    \node (image) [anchor=south west]{\includegraphics[width=5cm]{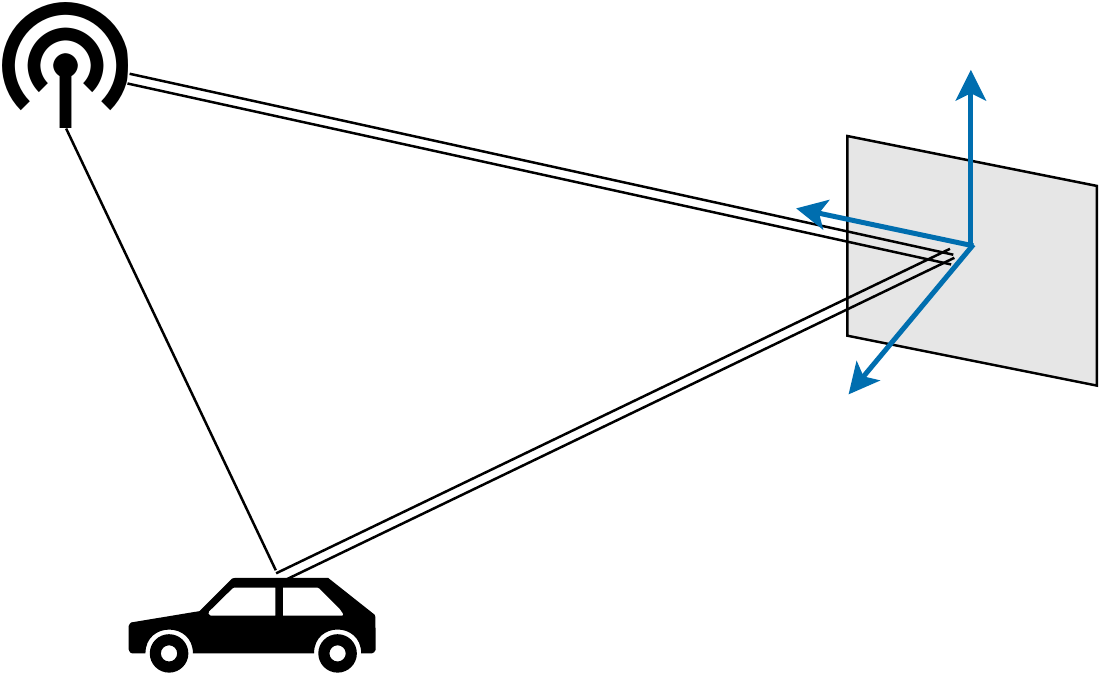}};
    \gettikzxy{(image.north east)}{\ix}{\iy};
    %\draw [help lines] (0,0) grid (\ix,\iy);
    \node at (2/5*\ix,.5/3.5*\iy){\footnotesize UE};
    \node at (.4/5*\ix,3.5/3.5*\iy){\footnotesize BS};
    \node at (4.5/5*\ix,1.3/3.5*\iy){\footnotesize RIS};
    %ue coordinates
    %\node at (1/5*\ix,.15/3.5*\iy){\footnotesize$\ueCoordinateVector{2}$};
    %\node at (.5/5*\ix,1/3.5*\iy){\footnotesize$\ueCoordinateVector{1}$};
    %\node at (1.3/5*\ix,1.7/3.5*\iy){\footnotesize$\ueCoordinateVector{3}$};
    %RIS coordinates
    \node at (4.3/5*\ix,3.3/3.5*\iy){\footnotesize$\vind{v}_3$};
    \node at (3.5/5*\ix,2.7/3.5*\iy){\footnotesize$\vind{v}_1$};
    \node at (4/5*\ix,1.4/3.5*\iy){\footnotesize$\vind{v}_2$};
    %positions
    \node at (4.5/5*\ix,2.1/3.5*\iy){\footnotesize$\risPosition$};
    \node at (1.5/5*\ix,.9/3.5*\iy){\footnotesize$\uePosition$};
    \node at (.7/5*\ix,2.7/3.5*\iy){\footnotesize$\bsPosition$};
    \node at (3.1/5*\ix,0/3.5*\iy){\footnotesize (a)};
    % right image
    \newcommand\w{3}
    \node (image2) at (\ix,0) [anchor=south west]{\includegraphics[width=\w cm]{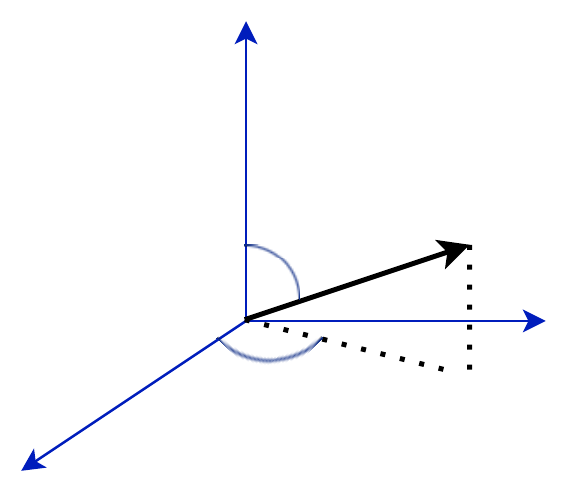}};
    \gettikzxy{(image2.north east)}{\ixt}{\iyt};
    %\draw [help lines] (\ix,0) grid (\ixt,\iyt);
    %coordinates
    \node at (\ix+.3/3*\w cm,.5/3*\iyt){\footnotesize$\vind{v}_1$};
    \node at (\ix+3/3*\w cm,.9/3*\iyt){\footnotesize$\vind{v}_2$};
    \node at (\ix+1.5/3*\w cm,3/3*\iyt){\footnotesize$\vind{v}_3$};
    %angels
    \node at (\ix+1.6/3*\w cm,.7/3*\iyt){\footnotesize$\psi_{\mr{az}}$};
    \node at (\ix+1.8/3*\w cm,1.5/3*\iyt){\footnotesize$\psi_{\mr{el}}$};
    \node at (\ix+.5*\w cm,0/3.5*\iy){\footnotesize (b)};
    \end{tikzpicture}
    \caption{(a): System setup, (b): Elevation and azimuth angles of a vector.}
    \label{fig:setup}
\end{figure}

We consider a wireless system consisting of a single-antenna BS, a single-antenna UE, and a RIS.  In Fig.\,\ref{fig:setup}a, the system setup is illustrated. The user receives the transmitted signal directly from  BS (LOS path) and also from the RIS (reflected path).
We assume that the position of the BS and the RIS ($\bsPosition$ and $ \risPosition$, respectively according to some general coordinate system) as well as the 
 rotation matrix corresponding to the RIS coordinate system 
\begin{align}
    \risRotationMatrix \defEq [\risCoordinateVector{1}, \risCoordinateVector{2}, \risCoordinateVector{3}]^\top\label{eq:risRotMat}
\end{align}
are known. On the other hand, 
the UE position ($\uePosition$) is unknown and should be estimated. In addition, we consider an asynchronous scenario, where the user clock bias $\timeBias$ should also be estimated. %as a nuisance parameter.
Fig.\,\ref{fig:setup}b presents the definition of azimuth and elevation of a general angle in the RIS coordinate system. 

We assume that RIS is an  $\Mrr$ by $\Mcc$ UPA with inter-element distance equal to $d$. In this case, the element in the $\ell$th row ($\ell \in \{0, \dots, \Mrr-1\}$) and $m$th column ($m \in \{0, \dots, \Mcc-1\}$) has the position $\risElementPosition{\ell,m} = [d\ell-d(\Mrr-1)/2), 0, dm-d (\Mcc-1)/2 ]$ in the RIS coordinate system. We assume that the RIS modulates the phase of the incident wave at time $t$ according to the matrix $\matInd{\Gamma}_t \in \mathbb{C}^{\Mrr\times \Mcc}$, where $\vert[\matInd{\Gamma}_t]_{\ell,m}\vert=1$ for all  $\ell$ and $m$. 

\subsection{Signal Transmission}\label{sec:SignalTransmission}
In this paper, we consider the transmission of $T$ OFDM symbols with $\numSubCarriers$ subcarriers for one localization occasion.  For simplicity, we assume that all the transmitted pilot symbols are equal to $\sqrt{\Es}$. Then the received signal can be expressed as the $\numSubCarriers\times T$ matrix
\begin{align}
&\matInd{Y} = \GainBs \sqrt{\Es} \vind{d}(\delayBS)\oneVector_{T}^\top 
+ \GainRis\sqrt{\Es} \vind{d}(\delayRis) \vind{u}(\aodRisToUe)^\top + \matInd{N}, \label{eq:channelModel}
\end{align}
where $\matInd{Y}=[\vind{y}_0,\dots, \vind{y}_{T-1} ]$, and $\vind{y}_t\in \complexSet^{\numSubCarriers}$ corresponds to the $t$th OFDM symbol. The noise matrix $\matInd{N}\in \complexSet^{\numSubCarriers\times T}$ comprises zero-mean circularly-symmetric independent and identically distributed  Gaussian elements with variance $\sigma^2$. The complex channel gain for the LOS path and the reflected path  are indicated by $\GainBs$ and $\GainRis$, respectively. The delays $\delayBS= {\vecNorm{\uePosition - \bsPosition}}/{c}+\timeBias$  and $\delayRis={\vecNorm{\uePosition - \risPosition}}/{c}+{\vecNorm{\bsPosition - \risPosition}}/{c}+\timeBias$ account for the propagation distance LOS and reflected path delays, respectively, including the clock offset $\Delta_t$. The effect of these delays on the signal is captured by the vectors $\vind{d}(\delayBS)$ and $\vind{d}(\delayRis)$, respectively, where we have defined
\begin{align}
    \vind{d}(\tau) &= [1, e^{-j2\pi \tau  \deltaF}, \dots, e^{-j2\pi \tau (\numSubCarriers-1) \deltaF}]^\top
\end{align}
with $\deltaF$ being the subcarrier spacing.

Furthermore, we have 
\begin{align}\nonumber
    [\vind{u}(\aodRisToUe)]_t &= \left(\risResponseVector{\aoaBsToRis}\right)^\top 
    \mr{diag}(\vind{\gamma}_t)\risResponseVector{\aodRisToUe}\\
    &= \left(\vind{\gamma}_t^\top\elementMultiply %
    \left(\risResponseVector{\aoaBsToRis}\right)^\top\right) \risResponseVector{\aodRisToUe}\label{eq:ur}
\end{align}
where, $\aoaBsToRis = [\aoaBsToRisAz,\aoaBsToRisEl]^\top$ represents the \emph{known} AOA in azimuth and elevation from the BS to the RIS, and $\aodRisToUe= [\aodRisToUeAz,\aodRisToUeEl]^\top$ denotes the AOD from the RIS to the UE, with $\aodRisToUeAz=\atant\left( \risToUeInRisCoorIndx{2}, \risToUeInRisCoorIndx{1} \right)$ and $\aodRisToUeEl =\arccos ({\risToUeInRisCoorIndx{3}}/{\vecNorm{\uePosition-\risPosition}})$, where $\risToUeInRisCoor = \risRotationMatrix (\uePosition-\risPosition)$.

Moreover, we have $\vind{\gamma}_t= \vect(\matInd{\Gamma}_t )$ and the response vector of the RIS for the angle $\generalAngle\in\{\aodRisToUe,\aoaBsToRis\}$ is defined as $\risResponseVector{\generalAngle} = \rispVecH{\generalAngle} \kron \rispVecT{\generalAngle}$ and 
\begin{align}
    \rispVecH{\generalAngle}&=e^{j \betar}[1, e^{-j [\waveNumberVector{\generalAngle}]_1d},\dots, e^{-j [\waveNumberVector{\generalAngle}]_{1} (\Mrr-1)d}]^\top\\
    \rispVecT{\generalAngle}&=e^{j \betac}[1, e^{-j [\waveNumberVector{\generalAngle}]_3d},\dots, e^{-j [\waveNumberVector{\generalAngle}]_{3} (\Mcc-1)d}]^\top
\end{align}
where $\betar = [\waveNumberVector{\generalAngle}]_{1}(\Mrr-1)d/2$ and $\betac = [\waveNumberVector{\generalAngle}]_{3}(\Mcc-1)d/2$. 
Here, 
\begin{align}
    \waveNumberVector{\generalAngle} = -\frac{2\pi}{\lambda}[\sin \generalAngleEl \cos \generalAngleAz ,  \sin \generalAngleEl \sin \generalAngleAz, \cos \generalAngleEl]^\top \label{eq:waveNumberVec}
\end{align}
is the wavenumber vector and $\lambda$ is the carrier wavelength. Here, we assume that the wavelength remains relatively constant within the transmission bandwidth.

\subsection{Goal}
Our goal is to estimate the UE location $\uePosition$ and its clock bias $\timeBias$ from the observation $\matInd{Y}$, given the RIS phase profiles, the transmitted signals, as well as location and orientation of the BS and RIS.

\section{Fisher Information Analysis}\label{sec:peb}
\subsection{FIM Derivation}

In this section, we calculate the PEB, which is a lower bound on the root mean square error (RMSE) of any unbiased estimator of the position upon  observing $\matInd{Y}$. The PEB can be calculated as 
\begin{align}
    \text{PEB} &= \sqrt{\tr\left(\left[\fisherInfpo^{-1}\right]_{1:3,1:3}\right)}\label{eq:peb}
\end{align}
where $\fisherInfpo$ is the FIM of positional parameters 
\begin{align}
    \parameterPo = [\uePosition^\top ,\timeBias ,\GainBsI, \GainBsR, \GainRisI, \GainRisR]^\top.
\end{align}
Here, the subindices $\mr{r}$ and $\mr{i}$ indicate the real and imaginary parts, respectively, of the path gains $\GainBs$ and $\GainRis$. We can calculate $\fisherInfpo$ based on the Fisher information of the channel parameters $\fisherInfoch$ as 
$\fisherInfpo = \jacob^\top \fisherInfoch \jacob$.
The Jacobian matrix $\jacob\in\realSet^{8\times 8}$ is defined as
\begin{align}
    \jacob_{\ell,s} = \frac{\rond [\parameterCh]_\ell}{\rond [\parameterPo]_s}\label{eq:jacobElements}
\end{align}
where 
$\parameterCh = [\delayBS,\delayRis, \aodRisToUeAz, \aodRisToUeEl,  \GainBsI, \GainBsR, \GainRisI, \GainRisR]^\top$
is the set of channel parameters. $\fisherInfoch$ can be written as
%
% \begin{align}
%     \left[\fisherInfoch\right]_{r,s} = \frac{2}{\sigma^2}\sum\limits_{t=0}^{T-1}\sum\limits_{n=0}^{\numSubCarriers-1}\realPart\left\{\frac{\rond [\meanSig]^{*}_{n,t}}{\rond [\parameterCh]_{r}}\frac{\rond [\meanSig]_{n,t}}{\rond [\parameterCh]_{s}}\right\}\label{eq:fimpo}
% \end{align}
\begin{align}
    \fisherInfoch = \frac{2}{\sigma^2}\sum\limits_{t=0}^{T-1}\sum\limits_{n=0}^{\numSubCarriers-1}\realPart\left\{\frac{\rond [\meanSig]_{n,t}}{\rond \parameterCh} \left(\frac{\rond [\meanSig]_{n,t}}{\rond \parameterCh}\right)^\herm\right\}\label{eq:fimpo}
\end{align}
where $\meanSig \in \complexSet^{\numSubCarriers\times T}$ is the noiseless part of the received signal in \eqref{eq:channelModel}. 
The matrices $\fisherInfoch$ and $\jacob$ needed to calculate the PEB are derived in Appendix~\ref{app:fimch} and Appendix~\ref{app:jacob}, respectively. 

The CRB bound can  be calculated also for $\timeBias$ and the channel parameters. As an example, the CRB bound for $\timeBias$ can be represented as 
\begin{align}
    \sqrt{\mathbb{E}[(\timeBias-\tilde{\Delta}_t)^2]}\geq \sqrt{[\fisherInfpo^{-1}]_{4,4}}
\end{align}
where $\tilde{\Delta}_t$ indicates the estimate of the true time bias $\timeBias$ and $\mathbb{E}$ represents the expectation over  noise.

\subsection{FIM Interpretation}
We observe that the FIM involves a sum of rank-2 matrices over $T$ transmissions and $N$ subcarriers. As the dimension of $\fisherInfoch$ is $8\times8$, it immediately follows that $TN\ge 4$  is a necessary condition for the problem to be identifiable. On the other hand, setting $N>1$ is required to be able to estimate any of the delays, while $T>1$ is needed to estimate the AODs. From \cite{fascista2019millimeter}, it is known that in order for the AOD to be identifiable, we need that the RIS phase profile vectors $\bm{\gamma}_t$'s at different transmission instances are not all parallel. 

%From a practical perspective, the FIM analysis shows that the system is neither over- nor under-determined and that all geometric parameters (two TOAs and AOD in azimuth and elevation) are all needed to solve the 3D localization and clock bias estimation problem. 
From a geometric point of view, it can be shown that the four channel geometric parameters (i.e., the two delays and the two AODs for azimuth and elevation) determine the four positional parameters (the 3D localization as the intersection between the AOD line and one-half of a two-sheet hyperboloid, and the clock bias). There is a one-to-one mapping between the channel geometric and the positional parameters, which is equivalent to the fact that the Jacobian $\jacob$ is an invertible matrix.

\section{Low-complexity Estimation}\label{sec:estimation}
In this section, we develop an estimator to determine the channel parameters in $\parameterCh$ based on the received signal $\matInd{Y}$, and then give an estimation of the UE position and clock bias based on the channel parameters. We first estimate $\delayBS$ and $\delayRis$ using IFFT of the received signal and then $\aodRisToUe$ based on the 2D-IFFT of the RIS phase profile matrices. At the end of each step, a \emph{quasi-Newton} algorithm is used to refine the estimation.

\subsection{Estimation of $\delayBS$}\label{sec:taub}
We estimate $\delayBS$ based on the sum of the columns of $\matInd{Y}$, $\ySumCol = \sum_t \vind{y}_t $, with 
\begin{align}
\ySumCol &=  \GainBs \sqrt{\Es} \vind{d}(\delayBS)\oneVector_{T}^\top\oneVector_{T} + \GainRis\sqrt{\Es} \vind{d}(\delayRis) \vind{u}(\aodRisToUe)^\top \oneVector_{T} +  \vind{N}\oneVector_{T}, \notag \\
 & \approx T\GainBs \sqrt{\Es} \vind{d}(\delayBS) + \sum_t \vind{n}_t
\end{align}
where $\vind{n}_t$ represents the $t$th column of matrix $\matInd{N}$.
This operation adds coherently the $T$ transmissions for the LOS path, whereas this does not happen for the signal reflected by the RIS. We assume that $T$ times the received power from the BS is much greater than the received power from  RIS, which is a relevant assumption, especially when both BS and UE are in the far-field of  RIS. 

Next, we  calculate the  IFFT of the vector  $\ySumCol$, that is $\ySumColBar = \matInd{F} \ySumCol$, where  
  $\matInd{F}\in \complexSet^{\NF\times\numSubCarriers}$ is the IFFT matrix with elements 
\begin{align}\label{eq:ifft}
    \matInd{F}_{\ell,m} = \frac{1}{N_F}e^{2j\pi \ell m/N_F}
\end{align}
where $\NF$ determines the length of the IFFT vector. 
 If 
 \begin{align}
     \delayBS = \ell/(N_F\deltaF)\label{eq:cnd}
 \end{align}
 for some  integer $\ell$,  the absolute value of the elements of $\ySumColBar$ will likely have a maximum around $\ell$. However, with probability one, this assumption does not hold. Therefore, to obtain a more accurate estimation of $\delayBS$ we induce some artificial delay $\delta$ by calculating $\ySumCol(\delta) \defEq \ySumCol \elementMultiply \vind{d}(\delta)$. 
 First, we set $\delta = 0 $ and calculate
 $
     \tilde{k} = \arg\max\limits_{k} |\bm{e}^\top_k\matInd{F}\ySumCol(0)|,%\label{eq:taubMaxDelta0}
 $
 where $\bm{e}_k$ is vector comprising all zeros, except with a 1 in the $k$th entry. 
 Next, we refine our estimation by calculating
 \begin{align}
     \tilde{\delta}= \arg\max\limits_{\delta\in [0, 1/(N_F\deltaF)]} \left|\bm{e}^\top_k\matInd{F}\ySumCol(\delta) \right|.\label{eq:taubMax}
\end{align}
Then $\delayBS $ can be estimated as
\begin{align}
    \tilde{\tau}_{\bsMark} = \tilde{k}/(N_F\deltaF)-\tilde{\delta}.\label{eq:tauBopt}
\end{align}
 Based on our numerical observation, the optimization over $\delta$ can be performed by a quasi-Newton method using $0$ as starting point. We note that the purpose of auxiliary parameter  $\delta$ is to add an artificial delay to the signal such that $\delayBS + \delta$ satisfies the condition \eqref{eq:cnd}, therefore, it is sufficient to constrain the range of $\delta$ in (\ref{eq:taubMax}) to $[0, 1/(N_F\deltaF)]$.

\subsection{Estimation of $\delayRis$}\label{sec:taur}
To estimate $\delayRis$, we first remove the BS  contribution from the received signal $\matInd{Y}$ by computing
\begin{align}
    \bm{Y}_\risMark & = \bm{Y} - \tilde{g}_{\bsMark} \vind{d}(\tilde{\tau}_\bsMark)\oneVector_T^\top \\
    & \approx \GainRis\sqrt{\Es} \vind{d}(\delayRis) \vind{u}(\aodRisToUe)^\top + \matInd{N},
\end{align}
where 
\begin{align}
    \tilde{g}_{\bsMark} = \frac{1}{T\numSubCarriers}\sum\limits_{n=0}^{\numSubCarriers-1} [\ySumCol\elementMultiply \vind{d}(-\tilde{\tau}_{\bsMark})]_n
\end{align}
Then, similarly as in Section\,\ref{sec:taub}, we first perform the optimization
\begin{align}
     \tilde{k} = \arg\max\limits_{k} \vecNorm{\bm{e}^\top_k(\bm{F}\bm{Y}_{\risMark})}\label{eq:taurMax1}
\end{align}
where $\bm{e}^\top_k(\bm{F}\bm{Y}_{\risMark})$ extracts the $k$-th row from the IFFT of $\bm{Y}_{\risMark}$. To refine the estimate, we calculate 
%
%\begin{align}
%     \tilde{k} = \arg\max\limits_{k} \vecNorm{\bar{\vind{y}}_{k}(0)}\label{eq:taurMax1}
%\end{align}
%where $\bar{\vind{y}}_k(\delta)$ is the $k$th row of the %matrix 
%\begin{align}
%    F\left(Y_{\risMark}\elementMultiply\vind{d}(\delta)\oneVector_T^\top\right).
%\end{align}
%Next, we calculate
\begin{align}
     \tilde{\delta} = \arg\max\limits_{\delta\in[0, 1/(N_F\deltaF)]} \vecNorm{\bm{e}^\top_k \bm{F}\left(\bm{Y}_{\risMark}\elementMultiply\vind{d}(\delta)\oneVector_T^\top\right)}.\label{eq:taurMax}
\end{align}
Finally similarly as in \eqref{eq:tauBopt}, we have $\tilde{\tau}_{\risMark} = \tilde{k}/(N_F\deltaF)-\tilde{\delta}$.

%Next, we estimate $\delayRis$, based on the vector $ \vind{y}_\risMark = Y_\risMark \vind{h}$, where $\vind{h} \in \complexSet^{T}$ is chosen to be the eigenvector corresponding to the largest eigenvalue of the matrix $Y_\risMark^\hermY_\risMark$. This choice of vector $\vind{h}$ will maximize $\vecNorm{\vind{y}_\risMark}$ among all vectors with the same norm as $\vecNorm{\vind{h}}$. Then $\delayRis$ can be estimated similarly as in Sec.~\ref{sec:taub} by replacing $\hat{\vind{y}}$ with $\vind{y}_\risMark$.

\subsection{Estimation of $\aodRisToUe$}
To estimate $\aodRisToUe$, we first remove the delay effects of the reflected path from the signal $\matInd{Y}_\risMark$ by calculating
\begin{align}
    \matInd{Y}_{\aodRisToUe}= \matInd{Y}_\risMark \elementMultiply \vind{d}(-\tilde{\tau}_\risMark)\oneVector_T^\top.
\end{align}
Next, we sum over all the rows of $\matInd{Y}_{\aodRisToUe}$ to obtain $\vind{y}_{\aodRisToUe}\in \complexSet^{T\times 1}$. One can see from \eqref{eq:channelModel} that 
\begin{align}
\vind{y}_{\aodRisToUe} & =  \matInd{Y}_{\aodRisToUe}^\top \oneVector_{\numSubCarriers}\\
& \approx       N\GainRis\sqrt{\Es}  \vind{u}(\aodRisToUe) + \bm{N}^\top\oneVector_{\numSubCarriers}. \label{eq:yu}
\end{align}
To estimate $\aodRisToUe$ based on $\vind{y}_{\aodRisToUe}$,
we use the properties of the Kronecker product and $\vect$ operator  \cite[Eq.~(520)]{MatCookBook}, and based on \eqref{eq:ur}, we obtain
\begin{align}
    [\vind{u}(\aodRisToUe)]_t = e^{-j(\betar+\betac)} \rispVecT{\aodRisToUe}^\top\left(\matInd{\Gamma}_t\elementMultiply \matInd{A}(\aoaBsToRis)\right)\rispVecH{\aodRisToUe}\label{eq:ut}
\end{align}
where $\matInd{A}(\aoaBsToRis) = \rispVecT{\aoaBsToRis} \rispVecH{\aoaBsToRis}^\top$. Motivated by \eqref{eq:ut}, we search for $\aodRisToUe$ through taking 2D IFFT of the known matrices $\matInd{\Gamma}_t\elementMultiply \matInd{A}(\aoaBsToRis)$ as  
\begin{align}\label{eq:2dfft}
    \bar{\matInd{\Gamma}}_t = \FT \left(\matInd{\Gamma}_t\elementMultiply \matInd{A}(\aoaBsToRis)\right) \FH^\top.
\end{align}
The matrices $\FT\in \complexSet^{\NFr\times \Mrr}$ and $\FH\in \complexSet^{\NFc\times \Mcc}$ are defined similarly as in \eqref{eq:ifft}, where $\NFr$ and $\NFc$ determine the dimensions of the 2D IFFT transforms. We note that matrices $\bar{\matInd{\Gamma}}_t$ do not depend on the received signal and can be calculated offline.

%-------------  figure---------------

%-------------  figure---------------
\begin{figure*}[t]
    \centering
    \begin{subfigure}[b]{0.32\textwidth}
\definecolor{mycolor1}{rgb}{0.00000,0.44700,0.74100}%
\definecolor{mycolor2}{rgb}{0.85000,0.32500,0.09800}%
\definecolor{mycolor3}{rgb}{0.92900,0.69400,0.12500}%
\definecolor{mycolor4}{rgb}{0.49400,0.18400,0.55600}%
\definecolor{mycolor5}{rgb}{0.85000,0.32500,0.09800}%

\pgfplotsset{every tick label/.append style={font=\footnotesize}}
\begin{tikzpicture}

\begin{axis}[%
width=4cm,
height=4cm,
at={(1.011in,0.642in)},
scale only axis,
xmin=0,
xmax=35,
xlabel style={font=\color{white!15!black}},
xlabel={\footnotesize $r(\mr{m})$},
ymode=log,
xmode=log,
ymin=0.03,
ymax=3,
yminorticks=true,
ylabel style={font=\color{white!15!black}},
ylabel={\footnotesize  Error ($\mr{m}$)},
axis background/.style={fill=white},
title style={font=\bfseries},
axis x line*=bottom,
axis y line*=left,
legend style={legend cell align=left, align=left, draw=white!15!black},
legend pos=north west
]
\addplot [color=mycolor3, line width=1.0pt]
  table[row sep=crcr]{%
1	0.0850535608651167\\
1.1304301330547	0.0784423831726795\\
1.27787228571806	0.0732907368695701\\
1.44454533797118	0.0670610710488466\\
1.63295757860631	0.0621157072369203\\
1.84594445285661	0.0587118988319367\\
2.08671123345428	0.0551674195343467\\
2.35888125728045	0.0518598785170914\\
2.66655045352778	0.0492932522566367\\
3.01434898397847	0.0481904721389353\\
3.40751092303208	0.0479668919023437\\
3.8519530261085	0.0488754938147826\\
4.35436377182428	0.0500236107560261\\
4.92230401795188	0.049778028378095\\
5.56432078594902	0.0544941025901468\\
6.29007588641937	0.0594484321899232\\
7.1104913212092	0.0657192668757071\\
8.0379136503188	0.0664467568443651\\
9.08629979721206	0.0760630894930794\\
10.2714270887373	0.0961193913996186\\
11.611130690583	0.115264603511445\\
13.1255720114712	0.133377674441251\\
14.8375421153464	0.16391310598585\\
16.7728047076557	0.215840849894099\\
18.9604838573757	0.276632428430053\\
21.4335022896747	0.397194337229899\\
24.2290768451452	0.503972906726077\\
27.38927856185	0.711032920945973\\
30.9616658089443	0.941331351006032\\
35	1.3393988356615\\
};
\addlegendentry{\footnotesize  CRB Pos.}

\addplot [color=mycolor3, only marks, mark=x, mark options={solid, mycolor3}]
  table[row sep=crcr]{%
1	0.0856483026729237\\
1.1304301330547	0.0809499433604825\\
1.27787228571806	0.073289478740171\\
1.44454533797118	0.0707232164338043\\
1.63295757860631	0.0619102680344471\\
1.84594445285661	0.0596113509147174\\
2.08671123345428	0.0562792274107363\\
2.35888125728045	0.0532490756103695\\
2.66655045352778	0.0509206637395241\\
3.01434898397847	0.0496327271934983\\
3.40751092303208	0.0505091059356251\\
3.8519530261085	0.0505886016403241\\
4.35436377182428	0.0517643596273159\\
4.92230401795188	0.050851260547657\\
5.56432078594902	0.0570044079190596\\
6.29007588641937	0.0620037156262782\\
7.1104913212092	0.0691192355963613\\
8.0379136503188	0.0718587946798254\\
9.08629979721206	0.0827380440523711\\
10.2714270887373	0.102428740152076\\
11.611130690583	0.125813132265082\\
13.1255720114712	0.146284511336349\\
14.8375421153464	0.185440050905637\\
16.7728047076557	0.250140962854278\\
18.9604838573757	0.327223010846783\\
21.4335022896747	0.495335628515308\\
24.2290768451452	0.65937752898754\\
27.38927856185	0.937275635065412\\
30.9616658089443	1.2392076409665\\
35	2.00430649057501\\
};
\addlegendentry{\footnotesize Est. Pos.}

\addplot [color=mycolor1, dashed, line width=1.0pt]
  table[row sep=crcr]{%
1	0.0763577008672708\\
1.1304301330547	0.0702880079367554\\
1.27787228571806	0.0655988702441545\\
1.44454533797118	0.0598033110446902\\
1.63295757860631	0.0550311306730752\\
1.84594445285661	0.0516558947110714\\
2.08671123345428	0.0483663358188663\\
2.35888125728045	0.0454434502444532\\
2.66655045352778	0.0429594355861715\\
3.01434898397847	0.0416497070753873\\
3.40751092303208	0.0415372673792057\\
3.8519530261085	0.0422720675742676\\
4.35436377182428	0.0433611305700958\\
4.92230401795188	0.0432038759859207\\
5.56432078594902	0.0474367615268769\\
6.29007588641937	0.0519749139905728\\
7.1104913212092	0.0579483426078664\\
8.0379136503188	0.0591276859251033\\
9.08629979721206	0.0682701681763779\\
10.2714270887373	0.0870445094206791\\
11.611130690583	0.105531054868711\\
13.1255720114712	0.123470478842637\\
14.8375421153464	0.153476344717692\\
16.7728047076557	0.204075126307509\\
18.9604838573757	0.263852863991311\\
21.4335022896747	0.38169190243366\\
24.2290768451452	0.487947573020312\\
27.38927856185	0.692470852832922\\
30.9616658089443	0.921748829949026\\
35	1.31691639466135\\
};
\addlegendentry{\footnotesize  CRB  $\Delta_t$}
\addplot [color=mycolor5, only marks, mark=+, mark options={solid, mycolor5}]
  table[row sep=crcr]{%
1	0.0766396825732088\\
1.1304301330547	0.0724548499194319\\
1.27787228571806	0.06547592684701\\
1.44454533797118	0.0630745673608689\\
1.63295757860631	0.0546843883223885\\
1.84594445285661	0.0523384525513354\\
2.08671123345428	0.0490700798957566\\
2.35888125728045	0.0465106997167115\\
2.66655045352778	0.04421582903569\\
3.01434898397847	0.0426409022114609\\
3.40751092303208	0.0437314492514347\\
3.8519530261085	0.0438013116111429\\
4.35436377182428	0.0447349054056184\\
4.92230401795188	0.0440080259034263\\
5.56432078594902	0.0495112622215554\\
6.29007588641937	0.0540469250506201\\
7.1104913212092	0.0608332245191233\\
8.0379136503188	0.0639007165170914\\
9.08629979721206	0.0741745572573437\\
10.2714270887373	0.092484415333246\\
11.611130690583	0.115024217895375\\
13.1255720114712	0.135226532064038\\
14.8375421153464	0.173262316279552\\
16.7728047076557	0.236293081303221\\
18.9604838573757	0.311752561920029\\
21.4335022896747	0.47546881779468\\
24.2290768451452	0.638064686336384\\
27.38927856185	0.912352846702413\\
30.9616658089443	1.21210813323948\\
35	1.96984439188391\\
};
\addlegendentry{\footnotesize  Est. $\Delta_t$}

\end{axis}

\end{tikzpicture}%
    \end{subfigure}
    %\hspace{1cm}
    \begin{subfigure}[b]{0.32\textwidth}
\definecolor{mycolor1}{rgb}{0.00000,0.44700,0.74100}%
\definecolor{mycolor2}{rgb}{0.85000,0.32500,0.09800}%
\definecolor{mycolor3}{rgb}{0.92900,0.69400,0.12500}%
\definecolor{mycolor4}{rgb}{0.49400,0.18400,0.55600}%
\definecolor{mycolor5}{rgb}{0.85000,0.32500,0.09800}%

\pgfplotsset{every tick label/.append style={font=\footnotesize}}
\begin{tikzpicture}

\begin{axis}[%
width=4cm,
height=4cm,
at={(1.011in,0.642in)},
scale only axis,
xmin=0,
xmax=35,
xlabel style={font=\color{white!15!black}},
xlabel={\footnotesize $r(\mr{m})$},
ymode=log,
xmode=log,
ymin=1e-05,
ymax=.05,
yminorticks=true,
ylabel style={font=\color{white!15!black}},
ylabel={\footnotesize  Error ($\mr{m}$)},
axis background/.style={fill=white},
title style={font=\bfseries},
axis x line*=bottom,
axis y line*=left,
legend style={legend cell align=left, align=left, draw=white!15!black},
legend pos=north west
]
\addplot [color=mycolor3, line width=1.0pt]
table[row sep=crcr]{%
1	5.4439007213244e-05\\
1.1304301330547	5.44886141099746e-05\\
1.27787228571806	5.45529119792126e-05\\
1.44454533797118	5.46333556213276e-05\\
1.63295757860631	5.47363805343221e-05\\
1.84594445285661	5.48693347454083e-05\\
2.08671123345428	5.50381962440936e-05\\
2.35888125728045	5.52533846866207e-05\\
2.66655045352778	5.55279884502845e-05\\
3.01434898397847	5.5877410609846e-05\\
3.40751092303208	5.6316382381332e-05\\
3.8519530261085	5.6875864903915e-05\\
4.35436377182428	5.75831428014745e-05\\
4.92230401795188	5.84736351819094e-05\\
5.56432078594902	5.95928287209288e-05\\
6.29007588641937	6.09935581194485e-05\\
7.1104913212092	6.27364529472664e-05\\
8.0379136503188	6.48962778557491e-05\\
9.08629979721206	6.75568363422529e-05\\
10.2714270887373	7.08115248545369e-05\\
11.611130690583	7.47620648743291e-05\\
13.1255720114712	7.95286806417022e-05\\
14.8375421153464	8.52303322024554e-05\\
16.7728047076557	9.20056590546879e-05\\
18.9604838573757	9.99935353952568e-05\\
21.4335022896747	0.000109362446366402\\
24.2290768451452	0.000120268522834683\\
27.38927856185	0.000132916355355945\\
30.9616658089443	0.000147505015460472\\
35	0.000164266065139132\\
};
\addlegendentry{\footnotesize CRB $\tau_{\mr{b}}$}

\addplot [color=mycolor3, only marks, mark=x, mark options={solid, mycolor3}]
  table[row sep=crcr]{%
1	6.20942069028624e-05\\
1.1304301330547	5.60987447360693e-05\\
1.27787228571806	5.40518356541884e-05\\
1.44454533797118	6.01510176761569e-05\\
1.63295757860631	5.56939942963768e-05\\
1.84594445285661	5.61438579912946e-05\\
2.08671123345428	5.59829879403117e-05\\
2.35888125728045	5.7564227064426e-05\\
2.66655045352778	6.017212222981e-05\\
3.01434898397847	5.50153259840683e-05\\
3.40751092303208	5.67849879411998e-05\\
3.8519530261085	5.74890895850038e-05\\
4.35436377182428	5.51246383418208e-05\\
4.92230401795188	6.07175293000155e-05\\
5.56432078594902	5.98970178989397e-05\\
6.29007588641937	6.23816254134743e-05\\
7.1104913212092	6.11869178938239e-05\\
8.0379136503188	6.41839961469535e-05\\
9.08629979721206	6.7689885708952e-05\\
10.2714270887373	7.10252963824371e-05\\
11.611130690583	7.50939276974186e-05\\
13.1255720114712	8.22824447848156e-05\\
14.8375421153464	8.67264361679177e-05\\
16.7728047076557	9.17916794176545e-05\\
18.9604838573757	0.000104346876312166\\
21.4335022896747	0.000113164612073641\\
24.2290768451452	0.000121369707673873\\
27.38927856185	0.000130996018719301\\
30.9616658089443	0.000138193834568644\\
35	0.000157457462127455\\
};
\addlegendentry{\footnotesize Est. $\tau_{\mr{b}}$}

\addplot [color=mycolor1, dashed, line width=1.0pt]
  table[row sep=crcr]{%
1	0.0065544857170206\\
1.1304301330547	0.00632966655980896\\
1.27787228571806	0.00601782493186703\\
1.44454533797118	0.00603909619949545\\
1.63295757860631	0.00629731503573464\\
1.84594445285661	0.00650552465195236\\
2.08671123345428	0.00643680672879576\\
2.35888125728045	0.00618069833105135\\
2.66655045352778	0.00615486509948764\\
3.01434898397847	0.00648880005409159\\
3.40751092303208	0.00644670745826865\\
3.8519530261085	0.00666330984840211\\
4.35436377182428	0.00672191968379492\\
4.92230401795188	0.0066369292296753\\
5.56432078594902	0.00715317222754501\\
6.29007588641937	0.00756144285313814\\
7.1104913212092	0.00788437949532216\\
8.0379136503188	0.00744286057410133\\
9.08629979721206	0.00789197897278285\\
10.2714270887373	0.00920289012858515\\
11.611130690583	0.00986601294789026\\
13.1255720114712	0.0100120388320459\\
14.8375421153464	0.0106088783408662\\
16.7728047076557	0.01194597555691\\
18.9604838573757	0.0129586510524991\\
21.4335022896747	0.0156980339338779\\
24.2290768451452	0.0162496047304001\\
27.38927856185	0.0188098950203985\\
30.9616658089443	0.0199098502649313\\
35	0.0228635535860505\\
};
\addlegendentry{\footnotesize CRB $\tau_{\mr{r}}$}

\addplot [color=mycolor5, only marks, mark=+, mark options={solid, mycolor5}]
  table[row sep=crcr]{%
1	0.00717477463778398\\
1.1304301330547	0.00671074320611371\\
1.27787228571806	0.00647457963242623\\
1.44454533797118	0.00628770092848264\\
1.63295757860631	0.00656376100670987\\
1.84594445285661	0.00682415480806752\\
2.08671123345428	0.00695356660550967\\
2.35888125728045	0.00655831058468231\\
2.66655045352778	0.0065472588426572\\
3.01434898397847	0.00695574543851743\\
3.40751092303208	0.00675916458614228\\
3.8519530261085	0.0068231307579652\\
4.35436377182428	0.00709587148039649\\
4.92230401795188	0.00696005120993836\\
5.56432078594902	0.00755505393099315\\
6.29007588641937	0.00802350857542815\\
7.1104913212092	0.00842214595213956\\
8.0379136503188	0.00807928951026404\\
9.08629979721206	0.00864070219307441\\
10.2714270887373	0.0100897083949191\\
11.611130690583	0.0109105661997159\\
13.1255720114712	0.0111595401770557\\
14.8375421153464	0.0123312692546423\\
16.7728047076557	0.0140168789681156\\
18.9604838573757	0.0156240025342187\\
21.4335022896747	0.0200787834864292\\
24.2290768451452	0.021486138213659\\
27.38927856185	0.0251868742887774\\
30.9616658089443	0.027417924063564\\
35	0.0349057979386392\\
};
\addlegendentry{\footnotesize Est. $\tau_{\mr{r}}$}
\legend{};
\end{axis}
\draw (5cm,4.75cm) ellipse (.1cm and .2cm);
\node [rotate = 0] at (5cm,5.05cm){\footnotesize $\tau_{\mr{r}}$};
\draw (5cm,2.55cm) ellipse (.1cm and .2cm);
\node [rotate = 0] at (5cm,2.85cm){\footnotesize $\tau_{\mr{b}}$};
\end{tikzpicture}%
    \end{subfigure}
    \begin{subfigure}[b]{0.32\textwidth}
\definecolor{mycolor1}{rgb}{0.00000,0.44700,0.74100}%
\definecolor{mycolor2}{rgb}{0.85000,0.32500,0.09800}%
\definecolor{mycolor3}{rgb}{0.92900,0.69400,0.12500}%
\definecolor{mycolor4}{rgb}{0.49400,0.18400,0.55600}%
\definecolor{mycolor5}{rgb}{0.85000,0.32500,0.09800}%

\pgfplotsset{every tick label/.append style={font=\footnotesize}}
\begin{tikzpicture}

\begin{axis}[%
width=4cm,
height=4cm,
at={(1.011in,0.642in)},
scale only axis,
xmin=0,
xmax=35,
xlabel style={font=\color{white!15!black}},
xlabel={\footnotesize $r(\mr{m})$},
ymode=log,
xmode=log,
ymin=.0004,
ymin=.0001,
ymax=.0035,
yminorticks=true,
ylabel style={font=\color{white!15!black}},
ylabel={\footnotesize  Error ($\mr{rad}$)},
axis background/.style={fill=white},
title style={font=\bfseries},
axis x line*=bottom,
axis y line*=left,
legend style={legend cell align=left, align=left, draw=white!15!black},
legend pos=north west
]
\addplot [color=mycolor3, line width=1.0pt]
  table[row sep=crcr]{%
1	0.000652236459108158\\
1.1304301330547	0.000595673704475149\\
1.27787228571806	0.000556464173801551\\
1.44454533797118	0.000490413356372736\\
1.63295757860631	0.000429931417780023\\
1.84594445285661	0.000388512395108358\\
2.08671123345428	0.000344608510879368\\
2.35888125728045	0.000306391629920273\\
2.66655045352778	0.000278811591792966\\
3.01434898397847	0.000238010771737107\\
3.40751092303208	0.000217517635207655\\
3.8519530261085	0.000198233492488492\\
4.35436377182428	0.000192067587608084\\
4.92230401795188	0.00017259224758678\\
5.56432078594902	0.000160698075449256\\
6.29007588641937	0.000155720011026773\\
7.1104913212092	0.000153238681469052\\
8.0379136503188	0.000135094647166108\\
9.08629979721206	0.0001405558596776\\
10.2714270887373	0.000134178307291954\\
11.611130690583	0.000139432004317464\\
13.1255720114712	0.000150609235282467\\
14.8375421153464	0.000145635231264002\\
16.7728047076557	0.000162713018036254\\
18.9604838573757	0.00017063638500268\\
21.4335022896747	0.000178534440498949\\
24.2290768451452	0.000200327436876783\\
27.38927856185	0.000218369142787127\\
30.9616658089443	0.000246855017599248\\
35	0.000268725762145892\\
};
\addlegendentry{\footnotesize CRB. $\aodRisToUeEl$}

\addplot [color=mycolor3, only marks, mark=x, mark options={solid, mycolor3}]
  table[row sep=crcr]{%
1	0.000629990998769821\\
1.1304301330547	0.000611401561373424\\
1.27787228571806	0.00055945174538189\\
1.44454533797118	0.000520519738759031\\
1.63295757860631	0.000416904648334189\\
1.84594445285661	0.000394953352689813\\
2.08671123345428	0.000327589138468252\\
2.35888125728045	0.000303775990761494\\
2.66655045352778	0.000275823209093915\\
3.01434898397847	0.000240644455970323\\
3.40751092303208	0.000222748527744284\\
3.8519530261085	0.00020133505209515\\
4.35436377182428	0.000188003265080395\\
4.92230401795188	0.000166631557887022\\
5.56432078594902	0.00016249996518153\\
6.29007588641937	0.000150868069662141\\
7.1104913212092	0.000151494840294597\\
8.0379136503188	0.000137506246934274\\
9.08629979721206	0.00014299442157669\\
10.2714270887373	0.000130934719158631\\
11.611130690583	0.000139334001549149\\
13.1255720114712	0.000151810341481401\\
14.8375421153464	0.00014970415065385\\
16.7728047076557	0.000164733307540947\\
18.9604838573757	0.000171022507865547\\
21.4335022896747	0.000177660254065676\\
24.2290768451452	0.000195747283846816\\
27.38927856185	0.000214599259186449\\
30.9616658089443	0.000245122680287368\\
35	0.00026656195905302\\
};
\addlegendentry{\footnotesize Est. $\aodRisToUeEl$}

\addplot [color=mycolor1,dashed, line width=1.0pt]
  table[row sep=crcr]{%
1	0.00669095117856136\\
1.1304301330547	0.00539166340894507\\
1.27787228571806	0.00442567219356763\\
1.44454533797118	0.00347980010558586\\
1.63295757860631	0.00271241908269534\\
1.84594445285661	0.00214825211854669\\
2.08671123345428	0.00171159202473253\\
2.35888125728045	0.00138678311480582\\
2.66655045352778	0.00109688284676065\\
3.01434898397847	0.000840185615183715\\
3.40751092303208	0.000717585880592361\\
3.8519530261085	0.000598469005249172\\
4.35436377182428	0.000520099571122438\\
4.92230401795188	0.000422261138192352\\
5.56432078594902	0.000371587298015668\\
6.29007588641937	0.000326251536206362\\
7.1104913212092	0.00030750455610271\\
8.0379136503188	0.000261526492423934\\
9.08629979721206	0.000243311679124082\\
10.2714270887373	0.000233699920729737\\
11.611130690583	0.00023091474831668\\
13.1255720114712	0.000224227587748852\\
14.8375421153464	0.000233581696772296\\
16.7728047076557	0.000249130896520315\\
18.9604838573757	0.000252161509741837\\
21.4335022896747	0.000265640573427979\\
24.2290768451452	0.000290965557987858\\
27.38927856185	0.00031614535607028\\
30.9616658089443	0.000363819453856782\\
35	0.0004025667812637\\
};
\addlegendentry{\footnotesize CRB $\aodRisToUeAz$}

\addplot [color=mycolor5, only marks, mark=+, mark options={solid, mycolor5}]
 table[row sep=crcr]{%
1	0.00651868367323596\\
1.1304301330547	0.00556109845723417\\
1.27787228571806	0.0045252565490467\\
1.44454533797118	0.00366740149697713\\
1.63295757860631	0.00263866507155333\\
1.84594445285661	0.00219923326561741\\
2.08671123345428	0.00163542127887435\\
2.35888125728045	0.00137824360594758\\
2.66655045352778	0.0010773212603302\\
3.01434898397847	0.000829207789647164\\
3.40751092303208	0.000726059495009429\\
3.8519530261085	0.000606737433853607\\
4.35436377182428	0.000508699664625377\\
4.92230401795188	0.000424169954521437\\
5.56432078594902	0.000361085021317299\\
6.29007588641937	0.000308134129549823\\
7.1104913212092	0.000311593151227338\\
8.0379136503188	0.000267847372386929\\
9.08629979721206	0.000241024832242842\\
10.2714270887373	0.000234324138460638\\
11.611130690583	0.000229055265064288\\
13.1255720114712	0.00022459857141498\\
14.8375421153464	0.000233032442883606\\
16.7728047076557	0.000252692734645518\\
18.9604838573757	0.000251002283815445\\
21.4335022896747	0.000278411631859565\\
24.2290768451452	0.000287068452542263\\
27.38927856185	0.000319960954627397\\
30.9616658089443	0.000355165617404138\\
35	0.00040524826976704\\
};
\addlegendentry{\footnotesize Est. $\aodRisToUeAz$}
\legend{};
\end{axis}
\draw (3.1cm,3.6cm) ellipse (.1cm and .2cm);
\node [rotate = 0] at (3.1cm,3.95cm){\footnotesize $\aodRisToUeEl$};
\draw (3.5cm,5.1cm) ellipse (.2cm and .1cm);
\node [rotate = 0] at (3.95cm,5.1cm){\footnotesize $\aodRisToUeAz$};
\end{tikzpicture}%
    \end{subfigure}  %
    
    \caption{Estimation error (markers) and the CRB bounds (lines) for  user position, time bias $\timeBias$, AOD elevation $\aodRisToUeEl$, AOD azimuth $\aodRisToUeAz$, LOS delay $\delayBS$, and reflected path delay $\delayRis$, along the path $[-r/\sqrt{2},r/\sqrt{2},-10]$, where $r\in [1 \ 35]$.  }
    \label{fig:results}
\end{figure*}
%--------------------------------

Now, lets assume that for some integers $0\leq \ell\leq \NFr-1$ and $0\leq m\leq \NFc-1$, we have 
\begin{align}
    [\waveNumberVector{\aodRisToUe}]_1 d &\equiv -2\pi \ell/\NFr\label{eq:mm1} \quad \mod 2\pi\\
    [\waveNumberVector{\aodRisToUe}]_3 d &\equiv -2\pi m/\NFc \quad \mod 2\pi.\label{eq:mm2}
\end{align}
Then, by comparing \eqref{eq:ut} and \eqref{eq:2dfft}, one can see that for some complex scalar $\xi$ we have 
\begin{align}\label{eq:utGt}
    \vind{\eta}_{\ell,m} \defEq \left[[\bar{\matInd{\Gamma}}_0]_{\ell,m},\dots, [\bar{\matInd{\Gamma}}_{T-1}]_{\ell,m}\right]^\top = \xi \vind{u}(\aodRisToUe).
\end{align}
Motivated by \eqref{eq:utGt} and \eqref{eq:yu}, we find $\tilde{\ell}$ and $\tilde{m}$ such that $[\tilde{\ell}, \tilde{m}] = \arg\min_{\ell,m} w_{\ell,m}$, where 
\begin{align}\label{eq:wlm}
    w_{\ell,m} \defEq \vecNorm{\vind{y}_{\aodRisToUe} - h\lefto(\vind{\eta}_{\ell,m}\righto)\vind{\eta}_{\ell,m}}^2
\end{align}
where function $h(\vind{\eta}_{\ell,m})$ returns a scalar that minimizes the RHS of \eqref{eq:wlm} for a given vector $\vind{\eta}_{\ell,m}$ and is defined as 
\begin{align}
    h\lefto(\vind{v}\righto) \defEq \frac{\vind{v}^\herm\vind{y}_{\aodRisToUe}}{\vind{v}^\herm \vind{v}}.
\end{align}
Next, we use a quadratic interpolation to obtain continuous (and more accurate) values for $\tilde{\ell}, \tilde{m}$, i.e., we calculate 
\begin{align}
    \ell_{\mr{q}} &= \tilde{\ell} + \frac{{w}_{\ell-1,m}-{w}_{\ell+1,m}}{2(w_{\ell-1,m}+{w}_{\ell+1,m}-2w_{\ell,m})}.\label{eq:elt}
\end{align}
Also, we calculate $m_{\mr{q}}$ similarly to \eqref{eq:elt}.
Next, based on $\ell_{\mr{q}}$ and $m_{\mr{q}}$, we estimate  $[\waveNumberVector{\aodRisToUe}]_1$ and $[\waveNumberVector{\aodRisToUe}]_3$ using \eqref{eq:mm1}--\eqref{eq:mm2} and calculate an estimation of $\aodRisToUe$ using the following relations
\begin{align}
[\waveNumberVector{\aodRisToUe}]_2&=-\sqrt{{4\pi^2}/{\lambda^2}-([\waveNumberVector{\aodRisToUe}]_1)^2-([\waveNumberVector{\aodRisToUe}]_3)^2}\\
    \aodRisToUeAz &= \mr{atan2}\left(-[\waveNumberVector{\aodRisToUe}]_2, -[\waveNumberVector{\aodRisToUe}]_1\right)\label{eq:estAz}\\
    \aodRisToUeEl &= \mr{acos}\left(-\lambda[\waveNumberVector{\aodRisToUe}]_3/2\pi\right).\label{eq:estEl}
\end{align}

Finally, we refine our estimation by performing the minimization
\begin{align}\label{eq:qn3}
    \tilde{\aodRisToUe} = \arg\min\limits_{\generalAngle} \vecNorm{\vind{y}_{\aodRisToUe} - h\lefto(\vind{u}(\generalAngle)\righto) \vind{u}(\generalAngle)}
\end{align}
where $\vind{u}(\generalAngle)$ is defined in \eqref{eq:ur}. We solve \eqref{eq:qn3} by applying the quasi-Newton algorithm, using \eqref{eq:estAz} and \eqref{eq:estEl} as the initial point.

%-----------table-----------------

\begin{table}[!t]
\vspace{.1cm}
	\caption{Parameters used in the simulation.}
	\label{table:par}
	\centering
	\begin{tabular}{l l l }
		\hline
		\hline
		Parameter&Symbol& Value\\
		\hline
		RIS dimensions & $\Mrr\times \Mcc$ &$64\times 64$\\
		Wavelength & $\lambda$ & $1 \ {\mathrm{cm}}$\\
		RIS element distance & $d$ &$0.5 \ {\mathrm{cm}}$\\
		Light speed & $c$ & $3\times 10^8 \ \mathrm{m/s}$\\
		Number of subcarriers & $\numSubCarriers$ & $3\, 000$\\
		Subcarrier bandwidth & $\deltaF$ & $120 \ \mr{kHz}$\\
		Number of transmissions & $T$ & $256$\\
		Transmission Power &$\numSubCarriers\Es$ & $20 \ \mathrm{dBm}$\\
		Noise PSD & $N_0$ & $-174 \ \mathrm{dBm/Hz}$\\
		UE's Noise figure& $n_f$ & $8 \ \mathrm{dB}$\\
		Noise variance& $\sigma^2=n_f N_0\deltaF$ & $-115.2 \ \mathrm{dBm}$\\
		BS position & $\bsPosition$ & $[5,5,0]$\\
		IFFT dimensions & $\NF$ & $4096$\\
		2-D IFFT dimensions & $ \NFr = \NFc$ & $256$\\
		\hline
		\hline
	\end{tabular}
\end{table}

%-------------

\subsection{Estimation of UE position and clock bias}
The UE position is calculated as $\tilde{\uePosition} =\risPosition + \tilde{\kappa}\risRotationMatrix^\top \waveNumberVector{\tilde{\aodRisToUe}} $, where $\waveNumberVector{\tilde{\aodRisToUe}} $ is defined in  \eqref{eq:waveNumberVec} and 
\begin{align}\label{eq:kappa_tilde}
    \tilde{\kappa} = &\arg\min\limits_\kappa \big(\lVert \kappa\risRotationMatrix^\top\waveNumberVector{\tilde{\aodRisToUe}}\rVert+\vecNorm{\bsPosition-\risPosition}\nonumber\\
    & \quad - \vecNorm{\bsPosition-\risPosition- \kappa\risRotationMatrix^\top\waveNumberVector{\tilde{\aodRisToUe}} }-(\tilde{\tau}_{\risMark}-\tilde{\tau}_{\bsMark})c\big)^2
\end{align}
where $c$ is the speed of the light. The value of $\tilde{\kappa}$ can be found numerically via a gradient-descent algorithm.
Then we estimate the clock bias as
\begin{align}\label{eq:dtcomp}
    \tilde{\Delta}_t = \tilde{\tau}_{\bsMark}-\vecNorm{\tilde{\uePosition}-\bsPosition}/c.
\end{align}

\emph{Complexity:} Finally, we note that the estimation of the channel parameters can also be done by searching over the parameters $[\delayBS,\delayRis, \aodRisToUeAz, \aodRisToUeEl]$ to maximize the likelihood function (or minimize the squared error function). However, this four-dimensional search would be much more complex than our method which has two one-dimensional and one two-dimensional search.

%\subsection{RIS codebooks}\label{sec:risCodeBooks}
%In this paper we study two RIS codebooks which are listed below.
%\begin{itemize}
 %   \item \emph{Random codebook:} All the RIS phase shifts $\risElementGains_{t,\ell,m}$ are drawn from a uniform distribution over $(0,2\pi]$, independently for all $t,\ell,m$.
 %   \item \emph{Zero-sum codebook:} The RIS phases are selected as
 %   \begin{align}
%        e^{j\risElementGains_{t,\ell,m}} = [\matInd{A(\aoaBsToRis)}]_{\ell,m} e^{j\tilde{\alpha}_{t,\ell,m}}\label{eq:ZS-CB}
%    \end{align}
%    where the first term in the RHS of \eqref{eq:ZS-CB} nullifies the effect of AOA from BS to the RIS, and the second term $\tilde{\alpha}_{t,\ell,m}$ is drawn uniformly from the set $\{0,\pi\}$, such that $\sum_t e^{j\tilde{\alpha}_{t,\ell,m}}=0$ for all $\ell,m$. This condition ensures that the contribution of the reflected path becomes zero in the vector  $\hat{\vind{y}} = \sum_t \vind{y}_t$ and the estimation of $\tau_b$ can be performed without any interference from the reflected path.
%\end{itemize}

%%--------------------------------------results----------
\section{Simulation Results}
In this section, we evaluate the CRB  presented in Sec.~\ref{sec:peb} and compare them to the RMSE of the estimator proposed in Sec.~\ref{sec:estimation}. We  study also the effect of the number of  RIS elements on  PEB. To evaluate the estimator performance, we average over $1\,000$ noise realizations. For each noise realization, $\timeBias$ is set to a random number uniformly drawn from the interval $[0, 1/\deltaF)$.  It is assumed that the RIS and the BS are at the same height, while UE is placed $10$ meters below them. The absolute coordinate system is set to be aligned with the RIS coordinates, i.e., $\risRotationMatrix$ is an identity matrix and $\risPosition = [0,0,0]$. All the phase shifts of the RIS elements are drawn from a uniform distribution over $[0,2\pi)$, independently. The amplitude of the channel gains $\GainBs$ and $\GainRis$ are calculated based on Friis' formula (using unit directivity for BS, UE, and  RIS elements) and their phase are set randomly between $[0, 2\pi)$. The rest of the parameters are presented in Table\,\ref{table:par}.

Fig.\,\ref{fig:results} illustrates the CRB and the estimation errors for the UE position, clock bias, and channel parameters along the direction $[-r/\sqrt{2},r/\sqrt{2},-10]$. 
As can be seen, positioning with submeter accuracy is possible for $r<30$. One can see that the proposed estimator is operating in close vicinity of the theoretical bounds. With low values of $r$, the UE is positioned almost beneath RIS and a relatively poor angle estimation is obtained. With $r>10 \mr{m}$ the estimation performance deteriorates with $r$ for all the parameters as the 
signal-to-noise ratio (SNR) decreases. As can be seen, the estimation error of $\timeBias$ is close to the PEB bound. This can be explained based on \eqref{eq:dtcomp} and the fact that the error in $\delayBS$ is much less than the position error. 

%------------figure--------------------

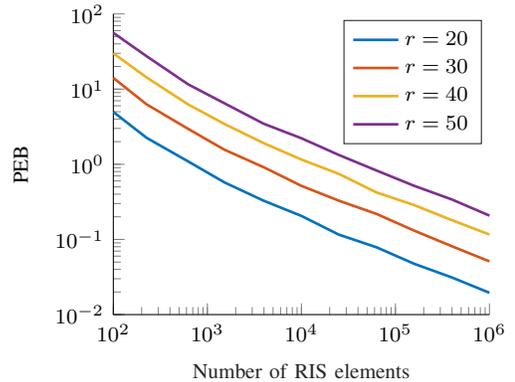
\begin{figure}[!t]
    \centering
\definecolor{mycolor1}{rgb}{0.00000,0.44700,0.74100}%
\definecolor{mycolor2}{rgb}{0.85000,0.32500,0.09800}%
\definecolor{mycolor3}{rgb}{0.92900,0.69400,0.12500}%
\definecolor{mycolor4}{rgb}{0.49400,0.18400,0.55600}%
\pgfplotsset{every tick label/.append style={font=\footnotesize}}

\begin{tikzpicture}

\begin{axis}[%
width=5cm,
height=4cm,
at={(1.011in,0.669in)},
scale only axis,
xmode=log,
xmin=100,
xmax=1000000,
xminorticks=true,
xlabel style={font=\color{white!15!black}},
xlabel={\footnotesize Number of RIS elements},
ymode=log,
ymin=0.01,
ymax=100,
yminorticks=true,
ylabel={\footnotesize PEB},
axis background/.style={fill=white},
axis x line*=bottom,
axis y line*=left,
legend style={legend cell align=left, align=left, draw=white!15!black}
]
\addplot [color=mycolor1, line width=1.0pt]
table[row sep=crcr]{%
100	4.97270498742053\\
225	2.25464893064859\\
625	1.09055507789486\\
1521	0.573634701782984\\
3969	0.32766306198312\\
10000	0.205579753674508\\
24964	0.115971567827134\\
63001	0.0785940798933476\\
158404	0.0474125125294374\\
396900	0.0312729296289702\\
1000000	0.0195210502329281\\
};
\addlegendentry{ \footnotesize $r = 20$}

\addplot [color=mycolor2, line width=1.0pt]
  table[row sep=crcr]{%
100	14.0959377584915\\
225	6.29040092659894\\
625	2.95726939751864\\
1521	1.56118861552971\\
3969	0.914378817690104\\
10000	0.516500173418493\\
24964	0.328246578927494\\
63001	0.218489898741255\\
158404	0.130930024549668\\
396900	0.0813948153170614\\
1000000	0.0510621830079995\\
};
\addlegendentry{ \footnotesize $r = 30$}

\addplot [color=mycolor3, line width=1.0pt]
  table[row sep=crcr]{%
100	29.9648719541795\\
225	14.3401264554063\\
625	6.25441440511451\\
1521	3.43651291904707\\
3969	1.91140306358336\\
10000	1.15698355947752\\
24964	0.745795942356189\\
63001	0.421633788110639\\
158404	0.286278322603045\\
396900	0.180956116465524\\
1000000	0.116298666613108\\
};
\addlegendentry{ \footnotesize $r = 40$}

\addplot [color=mycolor4, line width=1.0pt]
  table[row sep=crcr]{%
100	56.1008791490763\\
225	27.4714925250273\\
625	11.5755951834113\\
1521	6.47339675019026\\
3969	3.46366056162167\\
10000	2.2205229344728\\
24964	1.3320461509787\\
63001	0.825657531569633\\
158404	0.516814680821518\\
396900	0.340951182147777\\
1000000	0.207548485815688\\
};
\addlegendentry{ \footnotesize $r = 50$}

\end{axis}

\end{tikzpicture}%
    \caption{PEB for different numbers of RIS elements. Four different points are considered: $[-r/\sqrt{2}, r/\sqrt{2}, -10]$ with $r\in\{20,30,40,50\}$. }
    \label{fig:Mc}
\end{figure}
%--------------------------------

In Fig.\,\ref{fig:Mc}, we illustrate the effect of the RIS size on the PEB bound at five different locations.  A RIS with the same number of rows and columns is considered. One can see that for all the considered points, the decrease in the PEB is proportional to the square-root of the number of the RIS elements. This is due to the SNR increase of the reflected path, which is proportional to the number of RIS elements, and the fact that there is no significant beamforming gain because the phase shifts are random.

\section{Conclusion}
We have investigated localization and synchronization in a wireless system with a single-antenna UE, a single-antenna BS, and a RIS. We have calculated  the Cram\'er-Rao bounds and derived a low-complexity  estimator to determine the AOD from the RIS, and the delays of the direct and reflected signals. Then based on these parameters, we have estimated the user position and its clock bias. By comparing the estimator's error to the CRBs we have shown that our estimator is efficient and that 3D localization and synchronization is possible for the considered system. Our results point to the potential of RIS in enabling radio localization for simple wireless networks. 
A future research direction is to investigate the multi-user scenario and to study the effects of employing  multi-antenna BS,  different RIS phase profiles, and different pilots' energy levels on the localization performance in the considered setup. 
%--------------------
\section*{Acknowledgments}
This work was supported, in part, by the Swedish Research Council under grant 2018-03701, the Marie Sk\l{}odowska-Curie Individual Fellowships (H2020-MSCA-IF-2019) Grant 888913 (OTFS-RADCOM),
%the European Union’s Horizon 2020 research and innovation programme under the Marie Skłodowska-Curie grant agreement H2020-MSCA-IF-2017 798063,  --> Zohair's grant!
the Spanish Ministry of Science, Innovation and Universities under Projects TEC2017-89925-R and
PRX18/00638 and by the ICREA Academia Programme.

\appendices
\section{Calculating $\fisherInfoch$}\label{app:fimch}

In this appendix, we  derive $\rond[ \meanSig]_{n,t}/\rond [\parameterCh]_r$, then elements of $\fisherInfoch$ can be calculated from \eqref{eq:fimpo}. We have
\begin{align}
    \frac{\rond [ \meanSig]_{n,t}}{\rond \delayBS} &= -j2\pi n \deltaF  \GainBs \sqrt{\Es} e^{-j2\pi n \delayBS \deltaF} \\
    \frac{\rond [ \meanSig]_{n,t}}{\rond \delayRis} &= -j2\pi n \deltaF  \GainRis \sqrt{\Es} e^{-j2\pi n \delayRis \deltaF}[\vind{u}(\aodRisToUe)]_t\\
    \frac{\rond [ \meanSig]_{n,t}}{\rond\aodRisToUeAz} &= \GainRis \sqrt{\Es} e^{-j2\pi n \delayRis\deltaF}  \left(\frac{\rond \risResponseVector{\aodRisToUe}}{\rond \aodRisToUeAz}\right)^\top \vind{z}_t\\
    \frac{\rond [ \meanSig]_{n,t}}{\rond\aodRisToUeEl} &= \GainRis \sqrt{\Es} e^{-j2\pi n \delayRis\deltaF}  \left(\frac{\rond \risResponseVector{\aodRisToUe}}{\rond \aodRisToUeEl}\right)^\top \vind{z}_t
    \end{align}
where $\vind{z}_t= \vind{\gamma}_t\elementMultiply \risResponseVector{\aoaBsToRis}$ and
\begin{align}
    \frac{\rond\risResponseVector{\aodRisToUe}}{\rond \aodRisToUeAz} &= \risResponseVector{\aodRisToUe} \elementMultiply \left(-j(\frac{\rond\waveNumberVector{\aodRisToUe}}{\rond \aodRisToUeAz})^\top \matInd{Q}\right)\label{eq:rondAeq1}\\
    \frac{\rond\risResponseVector{\aodRisToUe}}{\rond \aodRisToUeEl} &=  \risResponseVector{\aodRisToUe} \elementMultiply \left(-j(\frac{\rond\waveNumberVector{\aodRisToUe}}{\rond \aodRisToUeEl})^\top \matInd{Q}\right)\label{eq:rondAeq2}\\
    \frac{\rond\waveNumberVector{\aodRisToUe}}{\rond \aodRisToUeAz} & = -\frac{2\pi}{\lambda_n}[-\sin \aodRisToUeEl \sin \aodRisToUeAz , \sin \aodRisToUeEl \cos \aodRisToUeAz , 0]^\top\label{eq:rondAeq3}\\
    \frac{\rond\waveNumberVector{\aodRisToUe}}{\rond \aodRisToUeEl} &= -\frac{2\pi}{\lambda_n}[\cos \aodRisToUeEl \cos \aodRisToUeAz , \cos\aodRisToUeEl \sin\aodRisToUeAz , -\sin \aodRisToUeEl]^\top\label{eq:rondAeq4}\\
    \matInd{Q} &= [\risElementPosition{0,0},\risElementPosition{1,0}, \dots, \risElementPosition{\Mrr-1,\Mcc-1}].\label{eq:rondAeq5}
\end{align}

 Finally, for gains, we have
\begin{align}
    [\frac{\rond [ \meanSig]_{n,t}}{\rond\GainBsR}, \frac{\rond [ \meanSig]_{n,t}}{\rond\GainBsI}]
    &= \sqrt{\Es}e^{-j2\pi n \delayBS \deltaF}[ 1,
      j]\\
      [ \frac{\rond [ \meanSig]_{n,t}}{\rond\GainRisR}, \frac{\rond [ \meanSig]_{n,t}}{\rond\GainRisI}]
    &= \sqrt{\Es}e^{-j2\pi n \delayRis \deltaF}[\vind{u}(\aodRisToUe)]_t[ 1,j].
\end{align}

\section{Calculating $\jacob$}\label{app:jacob}
According to \eqref{eq:jacobElements}, to determine the elements of $\jacob$, one should calculate the derivatives  ${\rond [\parameterCh]_{r}}/{\rond[\parameterPo]_{s}}$. To do so, we first write each element of $\parameterCh$ as a function of $\parameterPo$,  as described in Section \ref{sec:SignalTransmission} %in Appendix~\ref{app:function} 
and then calculate the derivatives.% in Appendix~\ref{app:jacobCalDrivatives}.
%\subsection{Expressing Channel parameters as a Function of Positional Parameters }\label{app:function}
%The elements of $\parameterCh$ can be determined based of $\parameterPo$ as described in Section \ref{sec:SignalTransmission}. 
%
%\begin{align}
%    \delayBS &= \frac{\vecNorm{\uePosition - \bsPosition}}{c}-\timeBias\\
    %
 %   \delayRis &= \frac{\vecNorm{\uePosition - \risPosition}}{c}+\frac{\vecNorm{\bsPosition - \risPosition}}{c}-\timeBias\\
    %
  %  \aodRisToUeAz &=\atant\left( \risToUeInRisCoorIndx{2}, \risToUeInRisCoorIndx{1} \right)\\
    %
   % \aodRisToUeEl &=\arccos \frac{\risToUeInRisCoorIndx{3}}{\vecNorm{\uePosition-\risPosition}}
%\end{align}
%where
%
%\begin{align}
%    \risToUeInRisCoor &= \risRotationMatrix (\uePosition-\risPosition)
%\end{align}
%where $\risRotationMatrix$ is defined in \eqref{eq:risRotMat}.

%\subsection{Calculating Derivatives}\label{app:jacobCalDrivatives}
%
For any function $f$, we use the notation $\rond f / \rond\uePosition \in \mathbb{R}^3$ to denote the gradient of $f$ with respect to variable $\uePosition$.

%[\rond f/\rond \uePosition_{1}, \rond f /\rond\uePosition_{2}, \rond f/\rond\uePosition_{3}]$. 
%
% tau
\begin{align}
\frac{\rond \delayBS}{\rond \uePosition}&= \frac{1}{c}\frac{\uePosition-\bsPosition}{\vecNorm{\uePosition-\bsPosition}}\label{eq:rond_tauBs}\\
\frac{\rond \delayRis}{\rond \uePosition}&= \frac{1}{c}\frac{\uePosition-\risPosition}{\vecNorm{\uePosition-\risPosition}}\\
[\frac{\rond \delayBS}{\rond \timeBias},\frac{\rond \delayRis}{\rond \timeBias}]&= [1, 1]. \label{eq:rond_tauRTb}
\end{align}
 For the AOD from RIS to UE, $\aodRisToUe$, we have
\begin{align}
        \frac{\rond\aodRisToUeAz}{\rond \uePosition}
    &= \frac{1}{ (\risToUeInRisCoorIndx{1})^2+(\risToUeInRisCoorIndx{2})^2} [-\risToUeInRisCoorIndx{2}\risRotationMatrix_{1,1:3}+\risToUeInRisCoorIndx{1}\risRotationMatrix_{2,1:3}]\\
    \frac{\rond\aodRisToUeEl}{\rond \uePosition}
&= \frac{-\vecNorm{\risToUeInRisCoor}^2[\risRotationMatrix]_{3,1:3}+(\uePosition-\risPosition)\risToUeInRisCoorIndx{3}}{\vecNorm{\risToUeInRisCoor}^2\sqrt{(\risToUeInRisCoorIndx{1})^2+(\risToUeInRisCoorIndx{2})^2}}.
\end{align}
 Furthermore, we have 
\begin{align}
    [ \frac{\rond\GainBsR}{\rond\GainBsR}, \frac{\rond\GainBsI}{\rond\GainBsI}, \frac{\rond\GainRisR}{\rond\GainRisR}, \frac{\rond\GainRisI}{\rond\GainRisI}] &= [1,1,1,1].
\end{align}
The rest of the derivatives are zero.
\if{0}
\fi
\balance

\end{document}